\documentclass[Journal]{IEEEtran}
\usepackage{amsmath,amsthm}
\usepackage{cite}
\usepackage{graphicx}
\usepackage{epstopdf}
\usepackage{color}
\usepackage{amsfonts,amsmath,amssymb}
\usepackage{cite}
\usepackage{graphicx}
\usepackage{url}
\usepackage{bm}
\usepackage{bbm}
\usepackage{amssymb}
\usepackage{bbold} %to create double ones for identity matrix
\def\BibTeX{{\rm B\kern-.05em{\sc i\kern-.025em b}\kern-.08em T\kern-.1667em\lower.7ex\hbox{E}\kern-.125emX}}
\usepackage{fancyhdr}
\usepackage{caption}
\usepackage{subcaption}
\DeclareCaptionFont{mysize}{\fontsize{8}{9.6}\selectfont}

\newcommand{\beq}{\begin{equation}}
\newcommand{\eeq}{\end{equation}}
\newcommand{\ket}[1]{| #1 \rangle}
\newcommand{\bra}[1]{\langle #1 |}
\newcommand{\ketbra}[3]{|#1 \rangle _{#3}\langle #2|}
\begin{document}
	\bstctlcite{IEEEexample:BSTcontrol}% force et al before citing any references

\title{Hybrid Entanglement Swapping for Satellite-based Quantum Communications}
\author{
    \IEEEauthorblockN{Hung Do\IEEEauthorrefmark{1}, Robert Malaney\IEEEauthorrefmark{1} and Jonathan Green\IEEEauthorrefmark{2}}\\
    \IEEEauthorblockA{\IEEEauthorrefmark{1}School of Electrical Engineering and Telecommunications,\\ The University of New South Wales, Sydney, NSW 2052, Australia.}\\
    \IEEEauthorblockA{\IEEEauthorrefmark{2}Northrop Grumman Corporation, San Diego, California, USA.}
}

\maketitle
\thispagestyle{fancy}
\renewcommand{\headrulewidth}{0pt}
%\vspace{-15mm}

\begin{abstract}
Hybrid entanglement swapping supports the teleportation of any arbitrary states, regardless of whether the quantum information in the state is encoded  in Discrete Variables (DV) or Continuous Variables (CV).  In this work, we study the CV teleportation channel created between two ground receivers via direct lossy-distribution from a low-Earth-orbit (LEO) satellite. Such a flexible teleportation protocol has the potential to interconnect a global array of quantum-enabled devices regardless of the different intrinsic technology upon which the devices are built.  However, past studies of hybrid entanglement swapping
have not accounted for channel transmission loss. Here we derive the general
framework for teleporting an arbitrary input mode over a lossy CV teleportation channel.
We investigate the specific case where the input modes are part of DV states entangled in the photon number basis, then identify the optimal teleportation strategy.
Our results show that, relative to DV photon-number entanglement sourced directly from the satellite, there are circumstances where our teleported DV states retain higher entanglement quality.
We discuss the implications of our new results in the context of generating a global network of ultra-secure communications between different quantum-enabled devices which possess line-of-sight connections to LEO satellites.
Specifically, we illustrate the impact the teleportation process has on the key rates from a Quantum Key Distribution protocol.
\end{abstract}

\section{Introduction}

Quantum communication via low-orbit satellites offers the possibility of an information-theoretically secure global network \cite{liao2017satellite}. The security of this network is provided by Quantum Key Distribution (QKD), in which secret keys are encoded and distributed between two parties using quantum states of light. Beyond QKD, other quantum-information protocols  will be relayed over this new global network, with teleportation being one of the most important \cite{BennetTele}.
%To eavesdrop on any encoded information, an eavesdropper must perform a quantum measurement on the quantum states. Such measurements will inevitably change the quantum states and thus be detected by the two legitimate parties during the reconciliation process. Following privacy amplification, the legitimate users then deliver a security key.

Terrestrial quantum communication  has been limited to a few hundred kilometers because of the exponential loss in optical fibers. What is more, according to the no-cloning theorem, the quantum signal cannot be noiselessly amplified \cite{park1970concept,dieks1982communication,wootters1982noclone}. A QKD experiment in 2007 achieved the communication range of 200km, which corresponds to 40dB of loss \cite{takesue2007QKDover40dB}, while a more recent study in 2015 used ultra-low-loss fiber to achieve a range of 300km with 50dB of total loss \cite{korzh2015provably}. In contrast, the satellite-Earth channel is only affected by a thin layer of atmospheric turbulence in a zone roughly between the ground level and 12 km high. For a satellite in low-orbit (400 to 600km from Earth), the average channel attenuation is only around 5-10dB for a down-link and 20-30dB for an up-link transmission \cite{peng2005experimental}. Thus, the satellite works as a relay that can potentially extend quantum communication to a global scale.

In many traditional quantum-information protocols, quantum entanglement is carried by either discrete variables (DV) or continuous variables (CV). In DV the quantum information is usually encoded into the polarization or the number of photons. In CV the quantum information is encoded in the quadratures of the optical field. Compared to DV  protocols, CV protocols use homodyne (or heterodyne) detectors which are faster and more efficient. As a result, it is potentially easier to achieve higher performance levels for certain CV quantum-information protocols.
%However, CV protocols only work well for high squeezing values, which makes it difficult for experimental realization \cite{he2018quantum, hosseinidehaj2018satellite}.
\begin{figure}[h!]
    \centering
        \includegraphics[width=\linewidth]{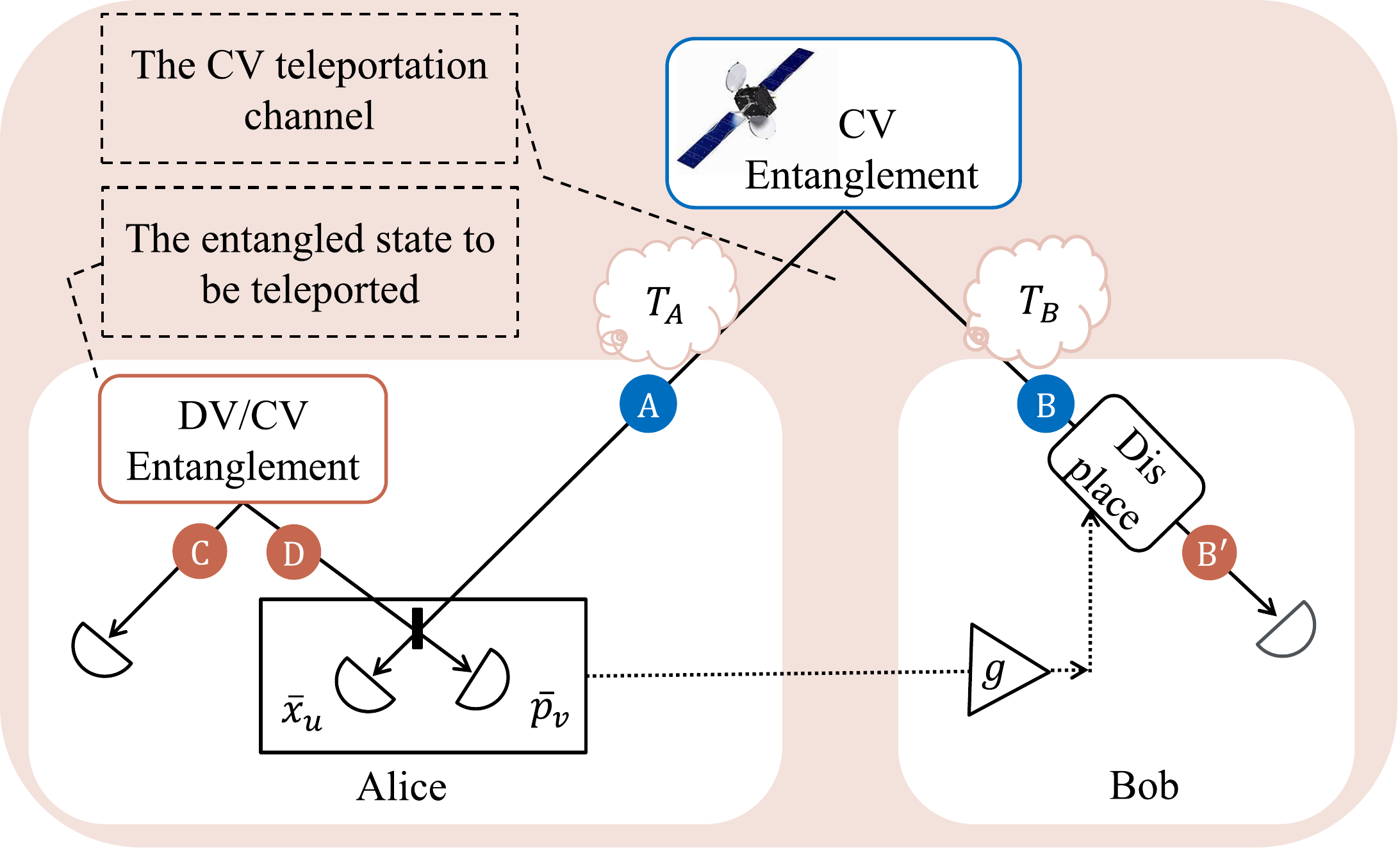}
		\caption{ In the hybrid scheme, the attenuated CV entanglement $A'-B'$ is used as a teleportation channel. The entangled state to be teleported, $C-D$, can be encoded in either DV or CV.}
		\label{setup}
\end{figure}

In this work we will investigate hybrid teleportation (Fig. \ref{setup}), where the CV entanglement is used as a teleportation channel ($A'-B'$), while the entangled state to be teleported ($C-D$) can be encoded in either DV or CV.
Such a hybrid teleportation protocol has the potential to leverage the versatility of the satellite. The design and implementation of a quantum satellite-payload often involves a great engineering effort. Once a satellite is deployed, we have little freedom in altering the payload and its functionality. However, if we place a source of CV entanglement on the satellite and beam the states down to the ground, we will have a CV teleportation channel that can teleport both DV and CV entangled states. This, in turn, opens up the possibility of inter-connecting quantum-enabled devices that may be processing quantum information internally on quite different hardware architectures - conditioned only on the premise that these devices have external laser links to  satellites. In this work we explore this possibility, focussing on the teleportation of DV states over a CV teleportation channel (CV teleportation via a CV teleportation channel has been well studied previously \cite{vaidman1994teleportation, braustein1998teleportationCV}).

 %In addition, hybrid DV-CV entanglement swapping leads to higher success rate as compared to DV-DV entanglement swapping \cite{takeda2015entanglementswapping}.%; the DV states after the swapping can also be postselected to achieve high-quality entanglement

Previous studies of entanglement swapping only focused on the context of local measurements without transmission loss \cite{vaidman1994teleportation,  braustein1998teleportationCV, braustein2005QIwithCV,liu2003improving, mista2010CVteleportation, takeda2013gaintuning, takeda2015entanglementswapping}. In this work, we extend such previous studies to the more difficult setting of photonic loss. Our main contributions and findings in this paper are:
\begin{itemize}

\item  We derive a mathematical model for hybrid teleportation over a CV teleportation channel, taking into account  transmission losses, and propose a strategy for gain-tuning in order to optimize the teleportation outcome.

\item We then perform a comparison between teleported and directly distributed DV entanglement (in the photon-number basis), showing that such  teleported DV entanglement can retain a higher quality, especially for an optical link loss of about 5 to 10dB.%, thus enabling  quantum-information protocols with improved performance (e.g., higher QKD key rates).

\item We calculate the key rates for a robust form of the device-independent QKD protocol, 
showing that key rate reduces to zero at around 2dB of loss. As a result, the teleported entanglement does not show a higher key rate relative to the directly-distributed entanglement.
%showing how it outperforms the key rate produced by directly distributed quantum states.

\end{itemize}

The structure of the remainder of this paper is as follows. Section \ref{attenuated_CV_channel} studies how channel transmissivities affect the quality of the CV teleportation channel. Section \ref{teleportation} studies hybrid entanglement swapping using an attenuated CV teleportation channel. Section \ref{directly_distributed_DV} studies the effect of channel attenuation on a directly-distributed DV entangled state.
%Section \ref{error_rate_key_rate} explains the calculation of the secure key rate for both teleported and directly-distributed DV entangled state.
 Section \ref{simulation} shows our simulations, while section \ref{conclusion} summarizes our findings.% some experimental challenges for satellite-QKD protocols using hybrid teleportation scheme.
%%%%%%%%%%%%%%%%%%%%%%%%%%%%%%%%%%%%%%%%%%%%%%%%%%%%%%%
%\input{CV_entanglement_20190801}

\section{Attenuation on a CV teleportation channel}
\label{attenuated_CV_channel}
%In this section, we introduce CV entanglement as a resource that facilitates teleportation over long distances (see illustration in Fig. \ref{setup}). The teleportation process is explained in the next section.
This section outlines how channel attenuation\footnote{Note, that in the downlink channel from the satellite some fading will exist. However, this can be mitigated by large receiving telescopes and/or  advanced adaptive optical tracking techniques, and in this exploratory work we will assume the channel loss is fixed (no fading).} affects the quality of the CV teleportation channel. Experimentally, CV entanglement can be created by combining two single-mode squeezed vacuum states through a balanced beam-splitter.\footnote{In this paper, we assume that there is no loss at this beam-splitter. When the loss is taken into account,  see  \cite{takeda2013gaintuning}.} Theoretically, the entangled state can be described by applying a two-mode squeezing operator on a two-mode vacuum state, thereby obtaining a two-mode squeezed vacuum (TMSV) state\cite{Girish2013QuantumOptics}. That is, $
\ket{\xi} = S(\xi)\ket{0,0} = \exp\left(\xi \hat{a}_A^\dagger \hat{a}_B^\dagger - \xi^* \hat{a}_A\hat{a}_B\right)\ket{0,0}$,
where $\xi = r e^{i\phi}$ and $r$ is the two-mode squeezing parameter. $\hat{a}$ and $\hat{a}^\dagger$ represent the annihilation and creation operators of the optical field, respectively, while the subscripts $A$ and $B$ denote the corresponding modes.

Let $\hbar =1 / 2 $, CV entanglement is encoded in the dimentionless quadrature variables of the optical field
\beq
\hat{x_i} = \frac{\hat{a_i}^\dagger+\hat{a_i}}{2}, \; \hat{p_i}=j\frac{\hat{a_i}^\dagger-\hat{a_i}}{2},
\eeq
respectively, where $i$ indicates the mode, and $j$ denotes the imaginary unit. The Wigner function of a TMSV state can then be written as \cite{braustein2005QIwithCV}
\begin{align}
&W_{TMSV}(x_A,p_A, x_B, p_B) = \frac{4}{\pi^2}\exp\{\nonumber\\
&-e^{-2r}\left[(x_A+x_B)^2+(p_A-p_B)^2\right]  \nonumber\\
&-e^{+2r}\left[(x_A-x_B)^2+(p_A+p_B)^2\right]\}.
\end{align}
We can see that when $r\rightarrow \infty$, this Wigner function approaches $C\delta(x_A - x_B)\delta(p_A + p_B)$, which represents a maximally entangled EPR state.

For each mode $i$ of the entangled state, the channel attenuation can be modeled by a beam-splitter with transmissivity $T_i$. The beam-splitter transformation can then be written as
\begin{align}
\hat{a_i}^{\dagger}  &\rightarrow \sqrt{T_i}\hat{a_i}^{\dagger} + \sqrt{1-T_i}\hat{a_i}^{\dagger}_{v},
\label{BStransformation}
\end{align}
where the subscript ${v}$ denotes the auxiliary vacuum mode. 
Let $G_{\sigma}(x,p)$ denote a Gaussian function with variance $\sigma$, where $x$ and $p$ are uncorrelated random variables with equal standard deviations ($\sqrt{\sigma}$)
\beq
G_\sigma(x,p) =\frac{1}{\pi\sigma}\iint{dx dp \exp\left(-\frac{x^2 + p^2}{\sigma}\right)}.
\label{Gaussian}
\eeq
In the Wigner-function formalism, each beam-splitter transformation is equivalent to a convolution between the input and a Gaussian function $G_{\sigma_{T_i}}$ with variance $\sigma_{T_i} = \frac{2T_i}{1-T_i}$, followed by a rescaling of $1/\sqrt{T_i}$ in phase space\cite{takeda2013gaintuning}. 
For a TMSV state, when the modes are sent through two channels with different transmissivities $T_A$ and $T_B$ (see Fig. \ref{setup}), we can perform the beam-splitter transformation sequentially on each mode: $W_{TMSV}(x_A, p_A, x_B, p_B) \rightarrow W_{TMSV}^{(T_A, 1)}(x_A',p_A', x_B, p_B) \rightarrow W_{TMSV}^{(T_A, T_B)}(x_A',p_A', x_B', p_B')$. Let $*$ denote the convolution operator, the first transformation (where $x_B$ and $p_B$ are taken as constants) is given by
\begin{align}
&W_{TMSV}^{(T_A, 1)}(x_A',p_A', x_B, p_B) = \frac{1}{T_A}\left[ W_{TMSV} \ast G_{\sigma_{T_A}}\right]  \left(\frac{x_A',p_A'}{\sqrt{T_A}}\right)\nonumber\\
&= \frac{1}{T_A}\iint dx_A\, dy_A W_{TMSV}(x_A,p_A, x_B, p_B)\times \frac{1}{\pi\sigma_{T_A}} \times\nonumber \\
& \;\; \exp\left\{-\frac{1}{\sigma_{T_A}}\left[\left(\frac{x_A'}{\sqrt{T_A}} - x_A\right)^2+\left(\frac{p_A'}{\sqrt{T_A}}-p_A\right)^2\right]\right\},
\end{align}
with the second transformation being similar, giving
%By simply writing $(\alpha_A', \alpha_B')$ as $(\alpha_A, \alpha_B)$, 
\begin{align}
&W_{TMSV}^{(T_A, T_B)}(x_A',p_A', x_B', p_B') = \frac{4}{\pi^2 \tau}\exp\{\nonumber\\
&-\frac{e^{-2r}}{\tau}\left[\left(x_A'\sqrt{T_B}+x_B'\sqrt{T_A}\right)^2+\left(p_A'\sqrt{T_B}-p_B'\sqrt{T_A}\right)^2\right]  \nonumber\\
&-\frac{e^{+2r}}{\tau}\left[\left(x_A'\sqrt{T_B}-x_B'\sqrt{T_A}\right)^2+\left(p_A'\sqrt{T_B}+p_B'\sqrt{T_A}\right)^2\right]\nonumber\\
&-\frac{2}{\tau}\left[(1-T_B)\left(x_A'^2+p_A'^2\right)
+(1-T_A)\left(x_B'^2 + p_B'^2\right)\right]\},
\label{TMSV_T}
\end{align}
where $\tau$ is given by
\beq
\tau = 1+\left[\cosh (2r)-1\right](T_A+T_B -2 T_A T_B).
\eeq

%%%%%%%%%%%%%%%%%%%%%%%%%%%%%%%%%%%%%%%%%%%%%%%%%%%%%%%%%%%%%%%%%%%%%%%

%\input{CV_teleportation}
\section{Teleportation by an attenuated CV teleportation channel}
\label{teleportation}
\subsection{Teleportation of an arbitrary mode}
\label{cv_teleportation}

We use the CV state in Eq. (\ref{TMSV_T}) to describe an attenuated CV teleportation channel ($A'-B'$ in Fig. \ref{setup}). The channel can teleport any modes, be it DV or CV, from Alice to Bob. Let $\alpha_i=x_i +jp_i$ where $i$ indicates the mode, the arbitrary input mode to be teleported (e.g. mode $D$) can be denoted as $W_{in}(\alpha_{in})$. The teleportation process starts with a Bell-state measurement (BSM).
%, where the input mode and mode $A$ are jointly measured and projected onto one of the four two-qubit Bell states 
\cite{vaidman1994teleportation, braustein1998teleportationCV, braustein2005QIwithCV}. Experimentally, this is done by combining the input mode with mode $A$ through a 50:50 beam splitter, which results in two new modes $u$ and $v$ such that
\begin{align}
&\hat{x}_u = \frac{\hat{x}_{in}-\hat{x}_A'}{\sqrt{2}},\;\;
\hat{p}_u = \frac{\hat{p}_{in}-\hat{p}_A'}{\sqrt{2}}\nonumber\\
&\hat{x}_v = \frac{\hat{x}_{in}+\hat{x}_A'}{\sqrt{2}},\;\;
\hat{p}_v = \frac{\hat{p}_{in}+\hat{p}_A'}{\sqrt{2}}.
\end{align}
We have $\alpha_A' = \left( \alpha_v - \alpha_u\right)/ \sqrt{2}$. After the beam splitter, the three-mode Wigner function is transformed to \cite{braustein2005QIwithCV}
\begin{align}
&W(\alpha_u,\alpha_v,\alpha_B') = \iint dx_{in} dp_{in}W_{in}(\alpha_{in})\nonumber\\
&\times W_{TMSV}^{(T_A, T_B)}\left(\frac{\alpha_v - \alpha_u}{\sqrt{2}},\alpha_B'\right)\nonumber\\
&\times \delta\left(\frac{x_u+x_v}{\sqrt{2}}-x_{in}\right)\delta\left(\frac{p_u+p_v}{\sqrt{2}}-p_{in}\right),
\end{align}
and Alice performs homodyne detections for $x_u$ and $p_v$. Alice's measurements of $x_u$ and $p_v$ are described by integration over $x_v$ and $p_u$, respectively. After a change of variable: $x_v \rightarrow x = (x_u+x_v)/\sqrt{2}$ and $p_u \rightarrow p = (p_u + p_v)/\sqrt{2}$, we have
\begin{align}
&\iint dx_v dp_u W(\alpha_u, \alpha_v, \alpha_B') = 2\iint{dx dp}W_{in}(x + jp)\nonumber\\
&\times W_{TMSV}^{(T_A, T_B)}\left[x-\sqrt{2}x_u +j\left( \sqrt{2}p_v-p\right), \alpha_B'\right].
\end{align}

After the BSM, Alice sends the result of $(x_u, p_v)$ to Bob who displaces his mode B from $\alpha_B'$ to $\alpha_B''$ such that $x_B' = (x_B'' - g\sqrt{2}x_u)$ and   $p_B' = (p_B'' - g\sqrt{2}p_v)$.
The parameter $g$ denotes the gain during the transformation, which can be experimentally tuned to optimize the output quality (see section \ref{simulation}). When Alice teleports an ensemble of input modes, the output should be averaged over all possible values of $x_u$ and $p_v$. The output of the whole teleportation process is 
\begin{align}
&W_{out}(\alpha_B'')= 2\iint{dx dp}W_{in}(x+jp) \nonumber\\
&\times \iint{dx_u dp_v}W_{TMSV}^{(T_A, T_B)}[x-\sqrt{2}x_u + j\left( \sqrt{2}p_v-p \right),\nonumber\\
&\qquad\qquad\qquad\qquad\quad x_B''-g\sqrt{2}x_u + j \left(p_B''-g\sqrt{2}p_v\right)].
\end{align}
After performing the integration over $x_u$ and $p_v$, we find that the output function is a Gaussian convolution of the input, followed by a rescaling of $\alpha_B'' \rightarrow \alpha_B''/g$ in phase space
\begin{align}
W_{out}(\alpha_B'') &= \frac{1}{g^2}\left[W_{in}*G_{\sigma_{tel}}\right]\left(\frac{\alpha_B''}{g}\right)\nonumber\\
&= \frac{1}{g^2}\iint dx dp W_{in}(\alpha) G_{\sigma_{tel}}\left(\frac{\alpha_B''}{g}-\alpha\right).
\label{general_teleportation}
\end{align}
Here $G_{\sigma_{tel}}$ is the Gaussian function defined in Eq. (\ref{Gaussian}) with variance $\sigma_{tel}$ given by
\begin{align}
\sigma_{tel} & = \frac{1}{4g^2}\;\,[\;\, e^{+2r}(g\sqrt{T_A}-\sqrt{T_B})^2 \nonumber\\
&\qquad\quad +e^{-2r}(g\sqrt{T_A}+\sqrt{T_B})^2 \nonumber\\
&\qquad\quad +2g^2\;\, (1-T_A)+2(1-T_B)\; ].
\label{sigma_general}
\end{align}
Eq. (\ref{sigma_general}) agrees with previous findings for simpler scenarios where there is no channel attenuation \cite{liu2003improving}, or where the gain is set to unity \cite{braustein2005QIwithCV}. 
When there is no channel attenuation, $T_A = T_B =1$, we have $
\sigma_{tel} = \left[e^{2r}(1-g)^2+e^{-2r}(1+g)^2\right]/{\left(4g^2\right)}$.
When the gain is set to unity, $g=1$, the variance is further simplified to $\sigma_{tel} = e^{-2r}$.
% so $W_{out}(\alpha_{out})= W_{in}*G_{e^{-2r}}(\alpha_{out})$. 

%%%%%%%%%%%%%%%%%%%%%%%%%%%%%%%%%%%%%%%%%%%%%%%%%%%%%%%%%%%%%%%%%%%%%%%%%%%
%\input{DV_CV_swapping_k}
%%%%%%%%%%%%%%%%%%%%%%%%%%%%%%%%%%%%%%%%%%%%%%%%%%%%%%%%%%%%%%%%%%%%
\subsection{Teleportation of a DV entangled state}
\label{hybrid_swapping}
In the previous subsection, we have discussed how to teleport an arbitrary input mode by an attenuated CV teleportation channel. In this section, we discuss the special case of hybrid DV-CV entanglement swapping (see Fig. \ref{setup}). In this scenario, the input mode $D$ is part of the DV entangled pair $C-D$. The teleportation is carried out by the CV teleportation channel $A-B$, which teleports the mode $D$ to $B''$. As a result, the teleportation output $\rho_{CB''}$ is a DV entangled state. In the Wigner formalism, the transformation of mode $D$ is described by Eqs. (\ref{general_teleportation}) to (\ref{sigma_general}). Since we are dealing here with a DV input, we will convert the above equations to the density matrix representation and provide a measure to qualify its level of entanglement.

In this work, we assume that the state $C-D$ is entangled in the photon-number basis. Experimentally, $\rho_{CD}$ is produced by passing a single photon through a balanced beam splitter, giving %\footnote{In this paper, we assume that the preparation of entanglement is perfect and the only loss comes from the transmission channel. For the cases when there is a loss in the entanglement preparation process, see \cite{takeda2013gaintuning}.}.
\begin{align}
	\ket{\psi}_{CD} &= (\ket{01}-\ket{10})/\sqrt{2} ,\quad
	\rho_{CD} = \ket{\psi}_{CD}\bra{\psi} \ .
	%&=\frac{1}{2}(\ketbra{0}{0}{C}\otimes\ketbra{1}{1}{D} +\ketbra{1}{1}{C}\otimes\ketbra{0}{0}{D} \nonumber\\
	%& \qquad+\ketbra{0}{1}{C}\otimes\ketbra{1}{0}{D} + \ketbra{1}{0}{C}\otimes\ketbra{0}{1}{D}).
\label{rhoCD}
\end{align}
In the Wigner-function representation, the teleportation will transform the subspace of mode $D$ according to \cite{takeda2013gaintuning}: $
W_{in}^{\ketbra{m}{n}{D}}(\alpha_D) \rightarrow \frac{1}{g^2}\left[W_{in}^{\ketbra{m}{n}{D}}(\alpha_D)*G_{\sigma_{tel}}\right]\left(\frac{\alpha_B''}{g}\right)$,
where $m,n \in \{0,1\}$ and $G_{\sigma_{tel}}$ is the Gaussian function of Eq. (\ref{Gaussian}), with variance $\sigma_{tel}$ defined in Eq. (\ref{sigma_general}). In the density-matrix representation%, the subspace of $B$ will be transformed to the  subspace of $B$ following the map $\ketbra{m}{n}{B} \rightarrow \hat{T}^{mn}_B $. We have
%\begin{align}
	%\rho_{CD}  \rightarrow\rho_{CB} & =  \frac{1}{2}(\;\;\,\ketbra{0}{0}{C}\otimes\hat{T}^{11}_{B}+\ketbra{1}{1}{C}\otimes\hat{T}^{00}_{B} \nonumber\\
	%& \quad\quad +\ketbra{0}{1}{C}\otimes\hat{T}^{10}_{B}+\ketbra{1}{0}{C}\otimes\hat{T}^{01}_{B}\;).
%\end{align}
, the output state can be decomposed to \cite{takeda2013gaintuning}
\begin{align}
\rho_{CB''} &= \sum_{k=-1}^{\infty}\rho_k, \,\textrm{where} \nonumber\\
\hat{\rho}_{k} &= a_{k} \ketbra{0}{0}{C}\otimes\ketbra{k}{k}{B''} \nonumber \\
&-b_{k}\ketbra{1}{0}{C}\otimes\ketbra{k}{k+1}{B''} -b_{k} \ketbra{0}{1}{C}\otimes\ketbra{k+1}{k}{B''}\nonumber\\
 &+c_{k} \ketbra{1}{1}{C}\otimes\ketbra{k+1}{k+1}{B''} , \,\textrm{where}
\label{hybrid_output}
\end{align}
\begin{align}
	a_k &= \frac{1}{2}T_{11\rightarrow kk} \,( k\geq 0),\text{or} \;0  \; (k = -1),\nonumber\\
		b_k & = \frac{1}{2}T_{10\rightarrow k+1 \,k}\, (k\geq 0), \text{or} \;0 \;(k=-1) ,\nonumber\\
		c_k & = \frac{1}{2}T_{00\rightarrow k+1 \, k+1} \,(k\geq -1), \,\textrm{and where}
		\label{abc}
\end{align}
\begin{align}
T_{00\rightarrow kk} &= \frac{2(\gamma-1)^k}{(\gamma+1)^{k+1}},\nonumber\\
T_{11\rightarrow kk} &= \frac{2(\gamma-1)^{k-1}}{(\gamma+1)^{k+2}}\left[(\gamma-2g^2+1)(\gamma-1)+4kg^2\right]\nonumber\\
T_{10\rightarrow k+1\,k} &= \frac{4g\sqrt{k+1}(\gamma-1)^k}{(\gamma+1)^{k+2}},
\label{T_transform}
\end{align}
with $\gamma \equiv g^2(2\sigma_{tel} + 1)$.
In the limit where $T_A = T_B =1$, $r\rightarrow \infty$ and $g=1$ (which leads to $\gamma = 1$), we have $a_1=b_0 = c_{-1} = 1/2$ while all other coefficients tend to 0. The output state $\rho_{CB''}$ becomes the same as $\rho_{CD}$ in Eq.  (\ref{rhoCD}), which is a maximally entangled state.
%From the output state $\rho_{CB}$ in Eq. (\ref{hybrid_output}), we can deduce the probability of n-photon pairs
%\begin{align}
%P(n) &= \sum_{k=-1}^{\infty} P_k(n), \textrm{   where}\nonumber \\
%P_k(n) &= \begin{cases}
%a_k & \textrm{  if  }  n=k \\
%c_k & \textrm{  if  } n= k+2 \\
%0 & \textrm{  otherwise}
%\end{cases}
%\label{Pn}
%\end{align}
%
%In reality, when the CV teleportation channel is attenuated ($T_A, T_B <1$), the teleportation output $\rho_{CB}$ contains not only vacuum states but also multiphoton states ($n>1$), which implies a decrease in the entanglement (see section \ref{simulation} for more details).

In order to assess the entanglement in $\rho_{CB''}$, we use  the logarithmic negativity: $0 \leq E_{LN} \leq 1$. When $E_{LN}=1$, $\rho_{CB''}$ is a pure entangled state. The logarithmic negativity of the teleported state (Eq. \ref{hybrid_output}) is \cite{takeda2013gaintuning}
\begin{align}
	E_{LN}(\rho_{CB''}) &= \log_2\left[1+\sum_{k=-1}^{\infty}\left(|\lambda^-_k|-\lambda^-_k\right)\right],
\label{E_LN_formula}
\end{align}
where	$\lambda^{\pm}_k$  are the eigenvalues of $\rho_k$
\beq
\lambda^{\pm}_k = \frac{1}{2} \left[a_k+c_k \pm \sqrt{(a_k - c_k)^2 + 4b^2_k}\right] .
\label{lambda_k}
\eeq

%%%%%%%%%%%%%%%%%%%%%%%%%%%%%%%%%%%%%%%%%%%%%%%%%%%%%%%%%%%%%%%%%%%%%%%%%%%
\begin{figure*}[h!]
    \centering
		\begin{subfigure}[b]{0.45\linewidth}
        \includegraphics[width=\linewidth]{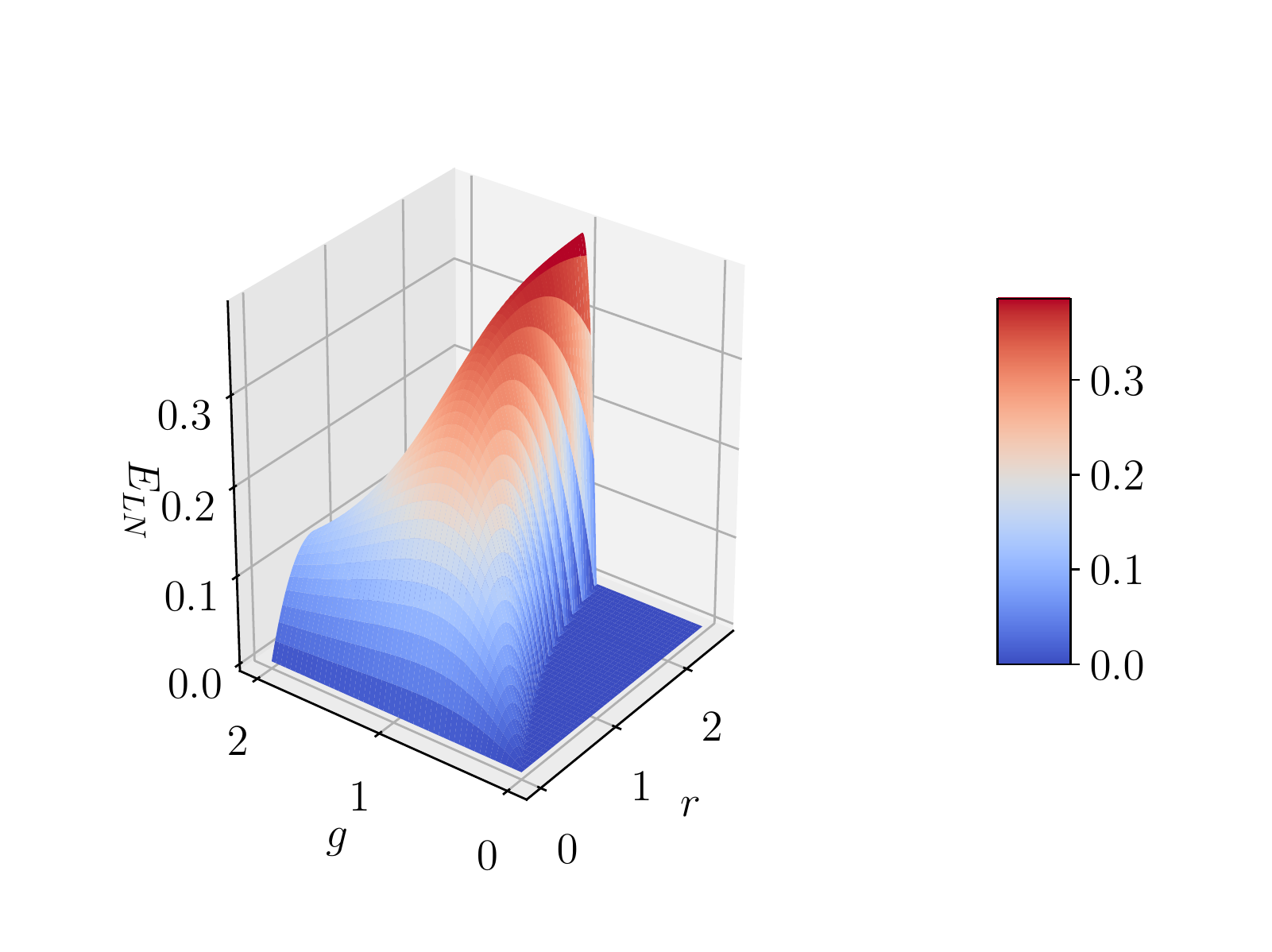}
        \caption{$E_{LN}$ when $T_A = T_B=0.562$.}
        \label{E_LN_r}
    \end{subfigure}
		\qquad
		\begin{subfigure}[b]{0.45\linewidth}
        \includegraphics[width=\linewidth]{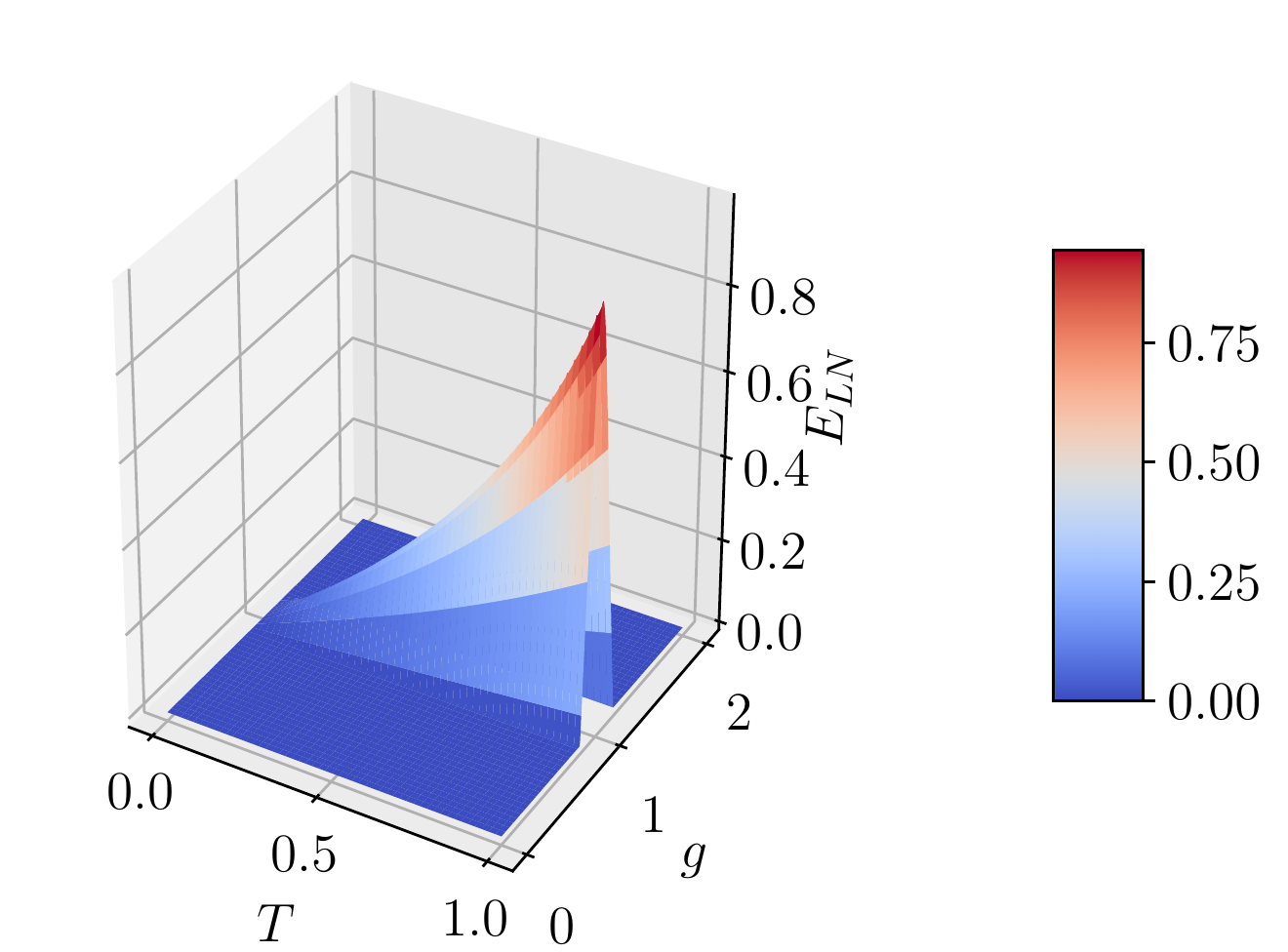}
        \caption{$E_{LN}$ when $T_A = T_B=T$.}
        \label{E_LN_T}
    \end{subfigure}
		
~
    \begin{subfigure}[b]{0.45\linewidth}
        \includegraphics[width=\linewidth]{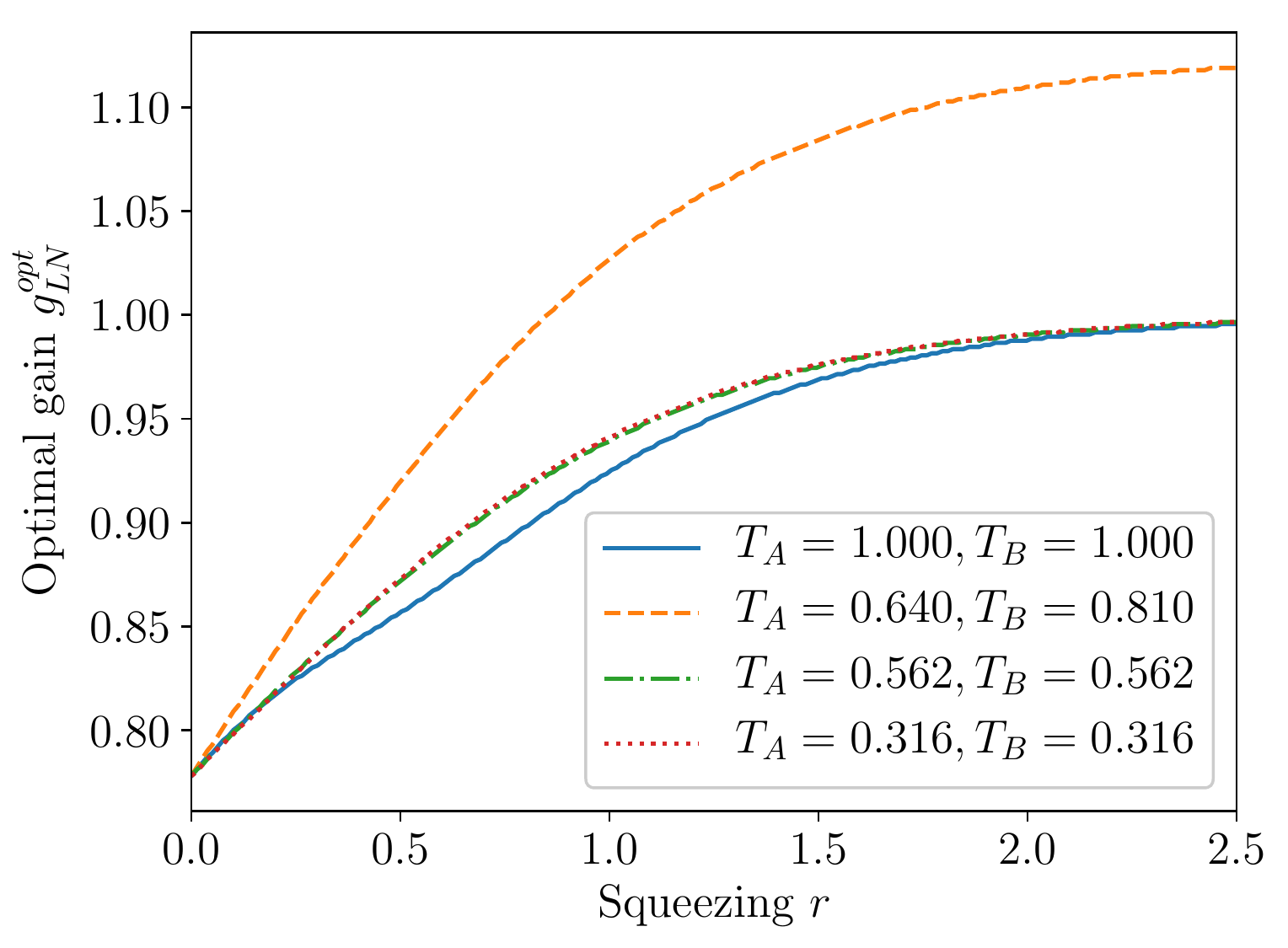}
        \caption{}
        \label{g_opt_r}
    \end{subfigure}
		\qquad
    \begin{subfigure}[b]{0.45\linewidth}
        \includegraphics[width=\linewidth]{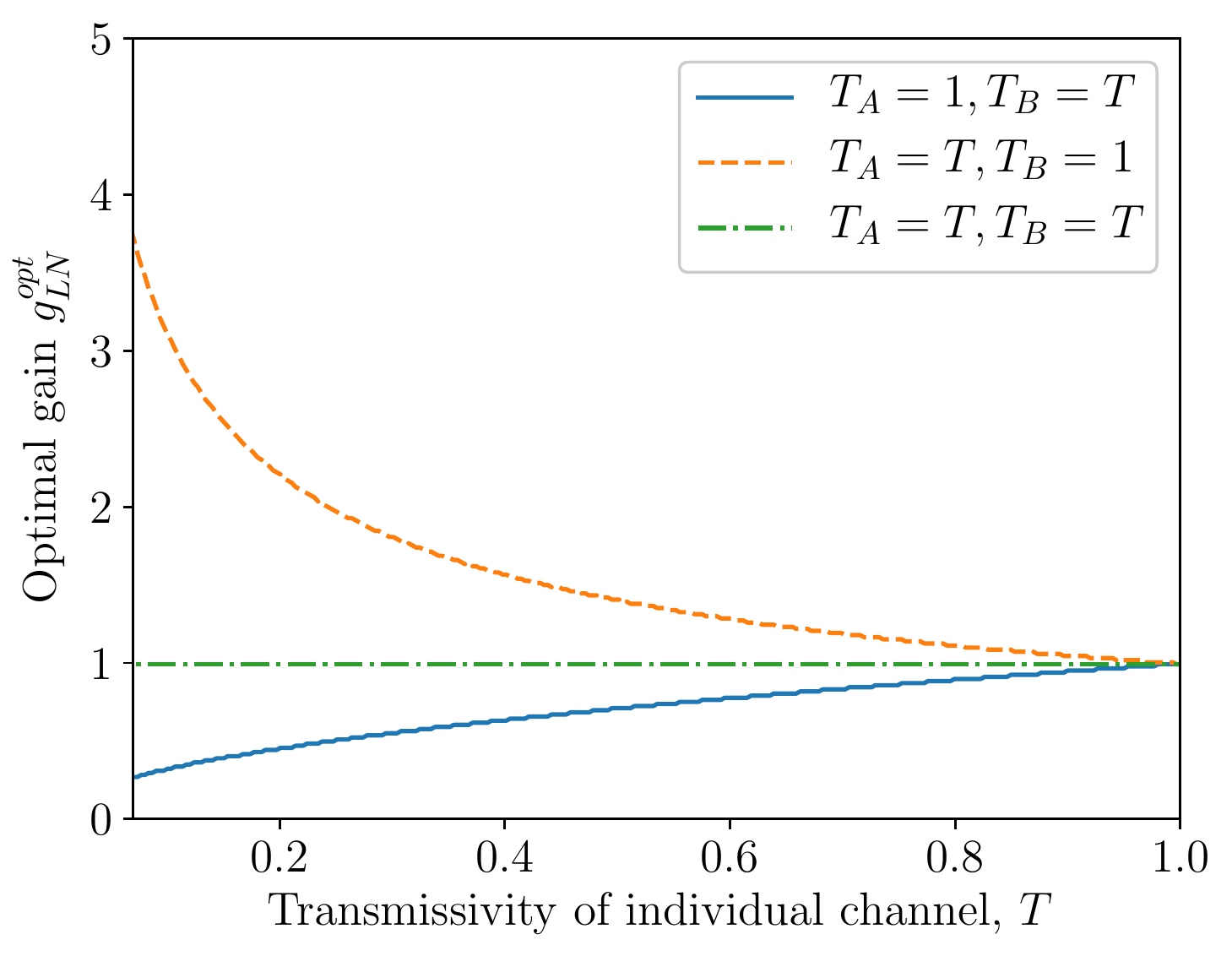}
        \caption{}
        \label{g_opt_T}
    \end{subfigure}
		\hfill
    ~ %add desired spacing between images, e. g. ~, \quad, \qquad, \hfill etc.

		\begin{subfigure}[b]{0.45\linewidth}
        \includegraphics[width=\linewidth]{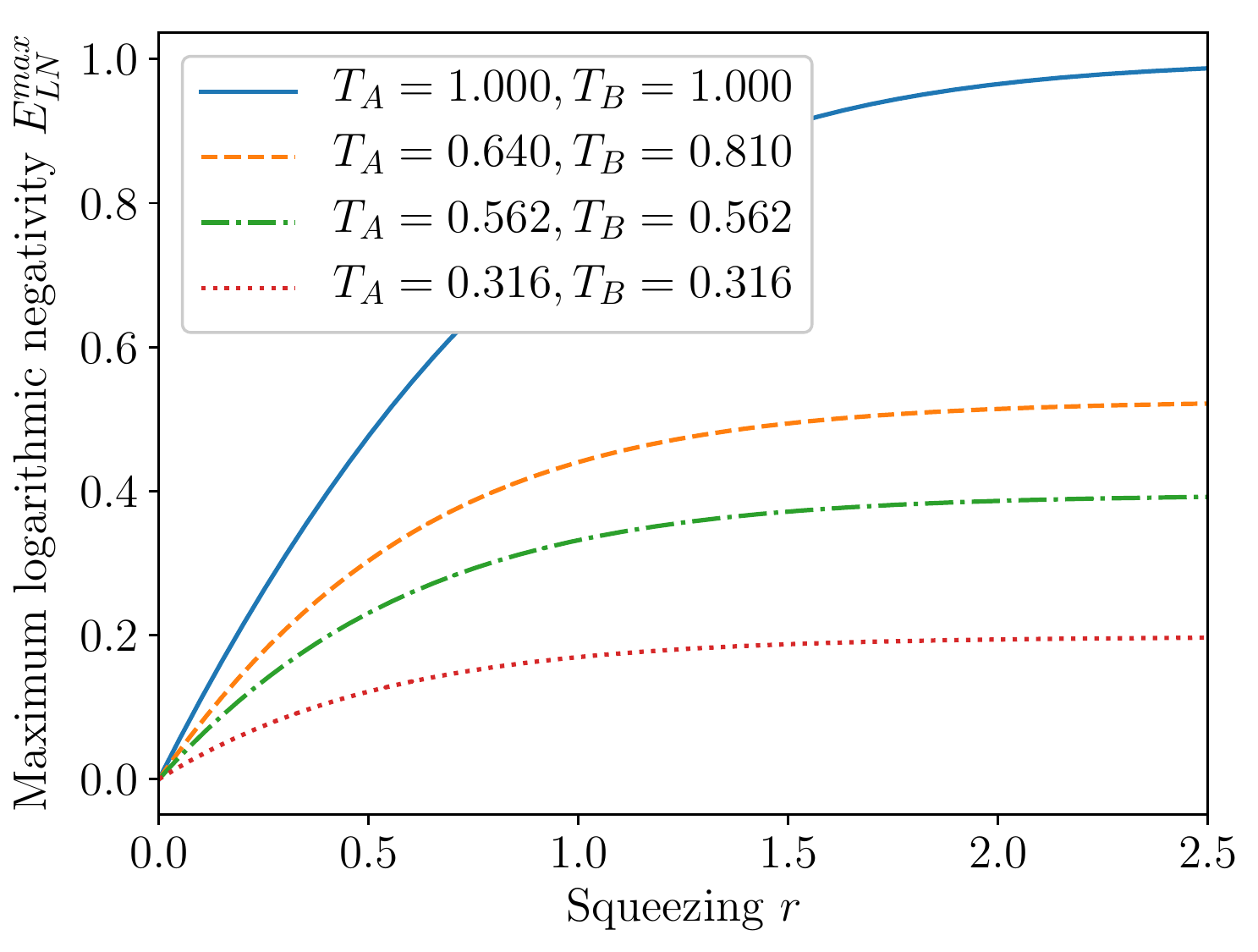}
        \caption{}
        \label{E_max_r}
    \end{subfigure}
		\qquad
    \begin{subfigure}[b]{0.45\linewidth}
        \includegraphics[width=\linewidth]{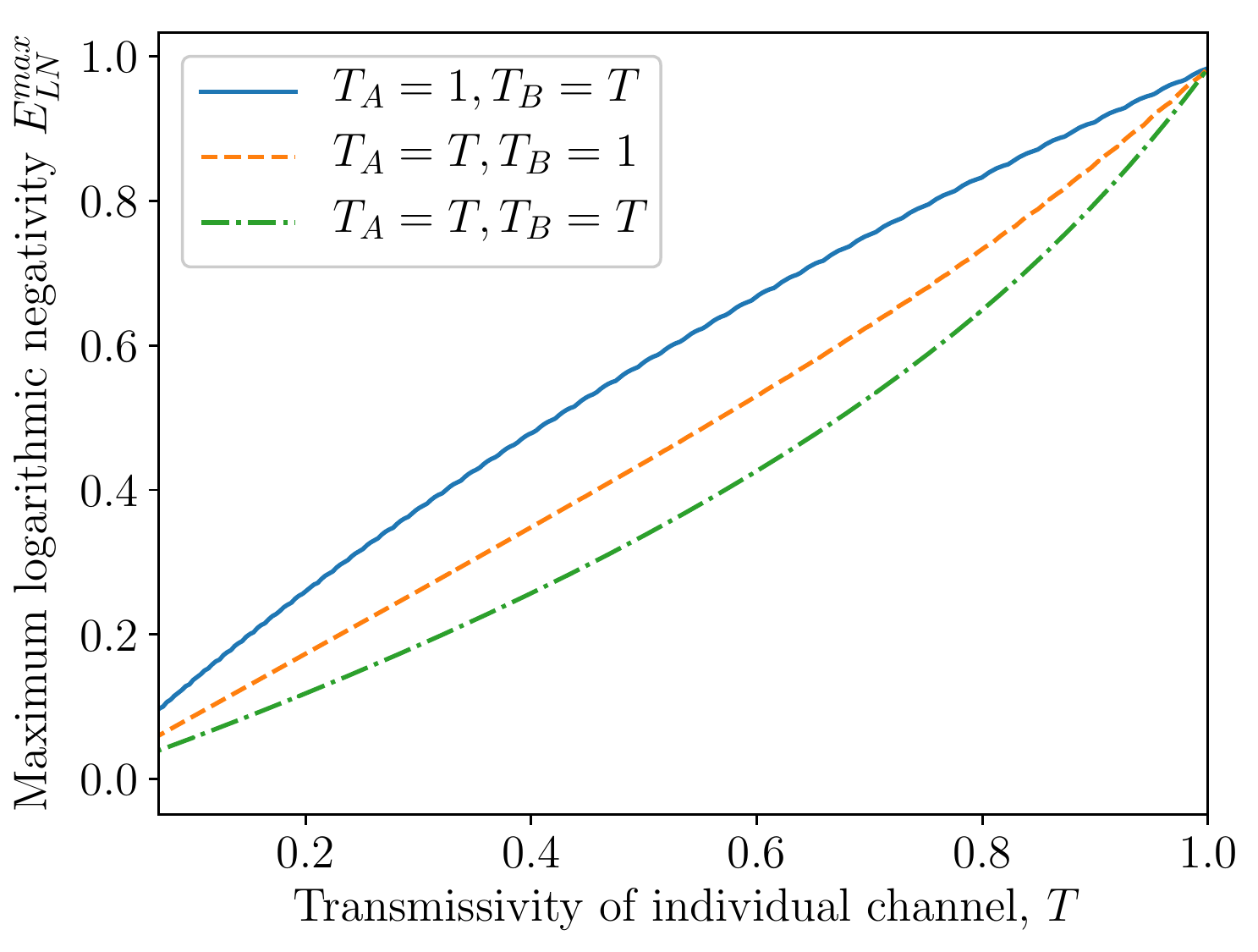}
        \caption{}
        \label{E_max_T}
    \end{subfigure}
    \caption{The logarithmic negativity $E_{LN}$ of the teleportation-output is plotted versus different values of squeezing $r$, channel transmissivities $T$ and teleportation gain $g$. In the left column, the squeezing ranges from 0 to 2.5 while the channel transmissivities are fixed at a few example values. It should be noted that when $T_A=T_B=0.562$ and $T_A=T_B=0.316$, the total channel loss is $-10\log_{10}\left(T_A T_B\right) = 5$dB and 10dB, respectively, which is the range of loss for a satellite down-link channel \cite{peng2005experimental}. In the right column, the squeezing is fixed at $r=2.395$ while the transmissivity $T$ ranges from 0 to 1. The first row shows the 3D plot of the logarithmic negativity when the gain $g$ is varied as well. The second row shows the optimal gain $g^{opt}_{LN}$ that maximizes the logarithmic negativity; while the third row shows the corresponding maximal logarithmic negativity $E_{LN}$. In general, we can see that $E_{LN}$ reaches its maximum value of 1 when the squeezing is high ($r>2.5$), when the gain is tuned to the optimal value $g^{opt}_{LN} = \sqrt{\frac{T_B}{T_A}}$, and when there is no channel loss, i.e., $T_A = T_B =1$. \\ \\ \\}
\label{g_opt_E_max}
\end{figure*}
%%%%%%%%%%%%%%%%%%%%%%%%%%%%%%%%%%%%%%%%%%%%%%%%%%%%%%%%%%%%%%%%%%%%%%%%%

%\input{directly_distributed_entanglement}
\section{Directly distributed DV entangled state}
\label{directly_distributed_DV}
In the last section, we have discussed how to teleport a DV-entangled state by a CV teleportation channel. In this section, we study the case when the DV entangled state ($A-B$) is directly distributed from the satellite. We also assume that the state is entangled in the photon-number basis\footnote{We note, conversion from DV-polarization coding to DV-photon-number encoding is possible, and it is part of  our future work to thoroughly study hybrid entanglement swapping in the context of attenuated CV teleportation  when the states to be teleported are encoded in polarization.},  and, for simplicity, that the two down-link channels have equal transmissivity of $T_A=T_B=T$. Let us denote the two modes of the state to be directly distributed as $A$ and $B$ with corresponding creation operators $\hat{a_A}^{\dagger}$ and $\hat{a_B}^{\dagger}$.  The state can be written as
\begin{align}
	\ket{\psi}_{AB} &= \frac{1}{\sqrt{2}}\left(\ket{01}_{AB}-\ket{10}_{AB}\right)\ket{00}_{{v}_A{v}_B}\nonumber\\
	&=\frac{1}{\sqrt{2}}(\hat{a_A}^{\dagger}-\hat{a_B}^{\dagger})\ket{00}_{AB}\ket{00}_{v_Av_B},
\end{align}
where $v_A$ and $v_B$ denotes the auxiliary vacuum modes of mode $A$ and $B$, respectively. For each mode of the entanglement, the channel attenuation can be modeled by a beam-splitter with transmissivity $T$. In the Heisenberg picture, the beam-splitter transformation can be written as in Eq. (\ref{BStransformation}), which transforms the state to
\begin{align}
\ket{\psi'}_{AB} &= \sqrt{\frac{1}{2}}(\sqrt{T}\hat{a_A}^{\dagger} + \sqrt{1-T}\hat{a_A}^{\dagger}_{v} \nonumber\\
	& \quad \; - \sqrt{T}\,\hat{a_B}^{\dagger} - \sqrt{1-T}\,\hat{a_B}^{\dagger}_{v} )\ket{00}_{AB}\ket{00}_{v_Av_B}\nonumber\\
	&= \sqrt{\frac{T}{2}} \left(\ket{01}_{AB} -\ket{10}_{AB} \right)\ket{00}_{v_A v_B} \nonumber\\
	& \quad \; + \sqrt{\frac{1-T}{2}} \ket{00}_{AB}\left(\ket{01}_{{v}_A{v}_B} -\ket{10}_{{v}_A{v}_B} \right).
\end{align}
By tracing out the auxiliary vacuum modes, we obtain
\beq
\rho_{AB}' = T \rho_{AB}+ (1-T)\ketbra{00}{00}{AB}.
\label{rho_direct}
\eeq
The logarithmic negativity of this state is calculated by using the relation
\beq
E_{LN}(\rho_{AB}') \equiv \log _2 ||\rho_{AB}^{'T_{trans}^A}||,
\label{E_LN_rho}
\eeq
where $T_{trans}^A$ denotes the partial transpose and
\beq
||\rho||= \textrm{Tr}\left[\left(\rho^\dagger\rho\right)^{1/2}\right].
\label{norm_rho}
\eeq 
%%%%%%%%%%%%%%%%%%%%%%%%%%%%%%%%%%%%%%%%%%%%%%%%%%%%%%%%%%%%%%%%%%%%%%%%%%
%\input{simulation_k}

\section{Simulation results}\label{simulation}

\subsection{Logarithmic negativity of teleported DV entanglement}
In this subsection, we determine the optimal gain $g^{opt}_{LN}$ and the corresponding maximal logarithmic negativity $E^{max}_{LN}$ achieved in different conditions. We simulate the teleportation process with different values of the squeezing parameter $r$, gain $g$ and the channel transmissivities $T_A, T_B$. The logarithmic negativity is calculated based on Eqs. (\ref{sigma_general}), (\ref{abc}), (\ref{T_transform}), (\ref{E_LN_formula}) and (\ref{lambda_k}). The results are shown in Fig. \ref{g_opt_E_max}. We first vary the squeezing $r$ while fixing the channel transmissivities at a few example values (Fig. \ref{g_opt_E_max}, left column). 
Next, we fix the squeezing at $r=2.395$ (so that $\cosh(2r) \approx 60$) while varying the channel transmissivities from 0 to 1 (Fig. \ref{g_opt_E_max}, right column). It can be seen that when the squeezing is high, the optimal gain tends to $\sqrt{\frac{T_B}{T_A}}$ (Fig. \ref{g_opt_E_max} (c),(d)). High squeezing also leads to higher maximum logarithmic negativities. In the case where $T_A = T_B =1$, $E^{max}_{LN}$ approaches 1 (Fig. \ref{g_opt_E_max} (e), (f)), and the output is a perfect copy of the input, which is a maximally DV entangled state.
%%%%%%%%%%%%%%%%%%%%%%%%%%%%%%%%%%%%%%

\subsection{Comparison between a teleported and a directly distributed DV entangled state}

\begin{figure}[h!]
    \centering
	%	\begin{subfigure}[b]{0.78\linewidth}
        \includegraphics[width=\linewidth]{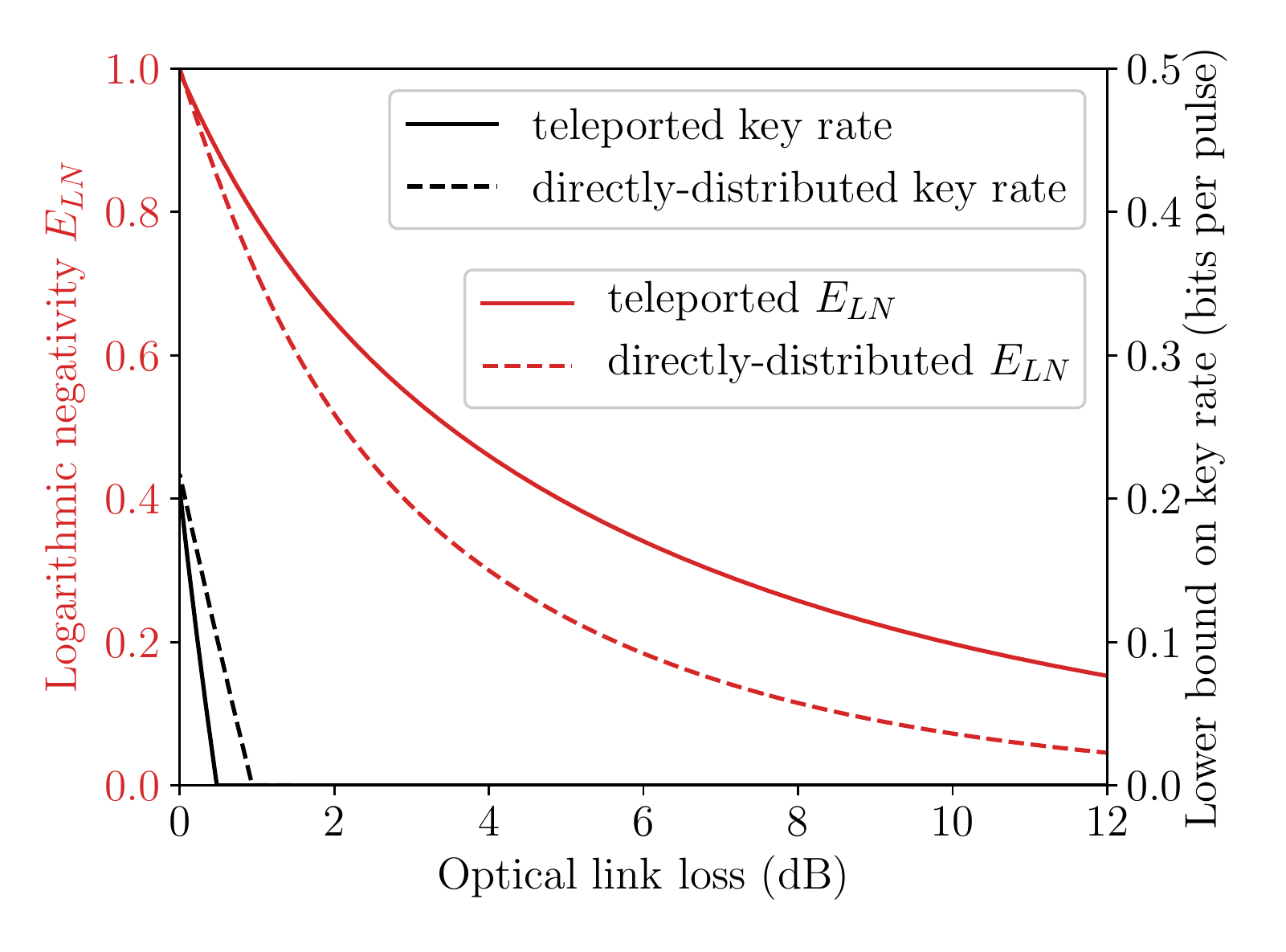}
    \caption{A comparison of DV entanglement distributed in two different ways: teleportation by a CV channel (solid lines) and direct down-link distribution from a satellite (dashed line). Here, the initial CV state (which formed the teleportation channel) had a squeezing value of $r=2.395$. The logarithmic negativity is plotted in red while the resulting bound on the secure key rate is plotted in black.}
		        \label{DV_sources_compare}
    %(a) shows the logarithmic negativity while (b) shows the produced secure key rate. Our finding shows that the former method retains higher logarithmic negativity and produces higher key rate. }
%\label{DV_compare}
\end{figure}
In this subsection, we compare a teleported DV entangled state with a DV entangled state directly distributed from a satellite. We assume that both states are encoded in the photon-number basis and are given by $\rho_{CB''}$ (Eq. (\ref{hybrid_output})) and $\rho'_{AB}$ (Eq. (\ref{rho_direct})), respectively. The channel is assumed to be symmetric, ($T_A=T_B=T$) so that the optimal gain is unity. The total optical link loss in dB is defined by $-10 \log_{10}(T^2)$.

The logarithmic negativity is calculated using Eqs. (\ref{E_LN_formula}), (\ref{lambda_k}), (\ref{E_LN_rho}) and (\ref{norm_rho}). At first, the squeezing value of the CV teleportation channel is set to a high value of $r=2.395$, so that $\cosh(2r)=60$. The result is plotted in red in Fig. \ref{DV_sources_compare}, where we can see that the teleported DV entangled state (solid line) retains higher logarithmic negativity than the directly-distributed DV entangled state (dashed line). We note that, when the optical link loss is from 5 to 10dB, the logarithmic negativity of teleported entanglement is more than double that of directly-distributed entanglement.

Next, we performed a numerical search to find the threshold squeezing $r_{th}$ where teleported entanglement starts to surpass directly-distributed entanglement. The average channel attenuation for the Earth-satellite channel is around 5 to 10dB for a down-link transmission\cite{peng2005experimental}. At this range of loss, as can be seen from Fig. \ref{DV_sources_compare}, the logarithmic negativity of directly-distributed entanglement is from 0.24 to 0.07. From Fig. \ref{g_opt_E_max} (e), specifically from the green dash-dotted line and the red dotted line, we can find that the teleported entanglement achieves the same level of logarithmic negativity at the threshold squeezing  $r_{th} = 0.3$ to $0.5$, %($\cosh(2r_{th})= 1.2$ to $1.6$), 
which is achievable experimentally. %In fact, when the squeezing is high enough ($r=2.394$), the logarithmic negativity teleported DV state has $E_{LN} = 0.4$ to $0.2$ while a directly distributed DV state has $E_{LN} = 0.2$ to $0.1$.

To calculate the lower bound of the secure key rates, we use a device-independent QKD (DI-QKD) protocol that exploits CH non-locality testing for DV entanglement in the photon-number basis. The security is analyzed against the individual attack where the eavesdropper is only constrained by no-signaling theory \cite{banaszek1999testing, kamaruddin2015DIQKD}. Our calculation follows closely reference \cite{kamaruddin2015DIQKD}, using the measurement setting of $s=0.5$.
%The only difference is that we optimize the key rate with two different arrangements of the CH inequality. For the directly-distributed state $\rho'_{AB}$, we use the same inequality in \cite{kamaruddin2015DIQKD}
%\begin{align}
%\langle CH_1 \rangle = &+Q_{AB}(0,0) + Q_{AB}(s,0) + Q_{AB}(0,-s)\nonumber\\
 %&- Q_{AB}(s,-s) - Q_A(0) - Q_B(0).
%\end{align}
%However, for the teleported state $\rho_{CB''}$, we rearrange the inequality to
%\begin{align}
%\langle CH_2 \rangle = &+Q_{AB}(0,0) + Q_{AB}(s,0)  + Q_{AB}(s,-s) \nonumber\\
%&- Q_{AB}(0,-s) - Q_A(s) - Q_B(0).
%\end{align}
We assume that the estimated bit error is zero and the squeezing is $r$=2.395. 
The result is plotted in black in Fig. \ref{DV_sources_compare}. The solid line represents the key rate bound from the teleported entanglement, while the dashed line represents the key rate bound from the directly-distributed entanglement. 
At zero loss, our simulations show that when the squeezing is  large ($r>5$), the key rate bound derived from the teleported state is approximately that of the key rate bound from the directly-distributed state.
When the loss increases, the key rate bounds become zero because the CH inequality is no longer violated. 
For our settings, the key rate bounds are reduced below zero before the optical link loss reaches 1dB. For this low loss, the teleported entanglement produces a lower key rate bound than the directly-distributed entanglement. However, we do note that this result is a consequence of the specific QKD protocol used and the fact that entanglement and CH inequalities are not always directly related. We anticipate the key rate behavior to be different for other QKD protocols.

\section{Conclusion}
\label{conclusion}

%In this work, we have studied the effect of channel transmission loss on the creation of a CV entanglement channel. We then use the CV teleportation channel to teleport DV entanglement in the photon number basis. 
%Our result shows that the entanglement in the teleportation output is maximized by increasing the squeezing above 2.3, tuning the gain to $g^{opt}_{LN} = \sqrt{\frac{T_B}{T_A}}$ and minimizing channel transmission loss.
%Somewhat surprisingly, we find that the teleported DV states can, at least for some parameter settings, retain a higher entanglement relative to the same states directly distributed from a satellite through lossy channels. Such a higher quality of the entanglement will lead to a performance improvement in many quantum-information protocols,  such as higher  key rates in  entanglement-based QKD protocols. We specifically determined such key rates for a very robust form of device-independent QKD.

In this work, we have studied the effect of channel transmission loss on the creation of a CV entanglement channel. We then use the CV teleportation channel to teleport DV entanglement in the photon number basis. 
Our result shows that the entanglement in the teleportation output is maximized by increasing the squeezing above 2.3, tuning the gain to $g^{opt}_{LN} = \sqrt{\frac{T_B}{T_A}}$ and minimizing channel transmission loss.
For our experimental settings, we find that the teleported DV states can retain a significantly higher entanglement relative to the same states directly distributed from a satellite, especially for an optical link loss from 5 to 10dB. For a device-independent QKD protocol, the minimum key rate reduces to zero at around 1dB of loss, before teleported entanglement shows a significant advantage over directly-distributed entanglement. 
Note, DI-QKD protocols are the most secure but are known to produce low key rates. 
%Our future work will seek a more suitable QKD protocol where the key rate remains above zero in the range of loss from 5 to 10 dB.

\bibliographystyle{IEEEtran}
\bibliography{myBib_short}

%\appendix
%\input{DIQKD}
\end{document}